\documentclass[english,aps, prl, twocolumn, nofootinbib, amsmath, amssymb, superscriptaddress, showpacs]{revtex4}
\usepackage[]{fontenc}
\usepackage[latin9]{inputenc}
\usepackage{amsmath,graphicx, mathtools}
\usepackage{pifont}
\usepackage{bbding}

\makeatletter
\@ifundefined{textcolor}{}
{%
 \definecolor{BLACK}{gray}{0}
 \definecolor{WHITE}{gray}{1}
 \definecolor{RED}{rgb}{1,0,0}
 \definecolor{GREEN}{rgb}{0,.4,0}
 \definecolor{BLUE}{rgb}{0,0,1}
 \definecolor{CYAN}{cmyk}{1,0,0,0}
 \definecolor{MAGENTA}{cmyk}{0,1,0,0}
 \definecolor{YELLOW}{cmyk}{0,0,1,0}
 }

\makeatother

\usepackage{babel}

\newcommand{\brac}[1]{\left( #1 \right)}

\newcommand {\ket}[1] {\lvert #1 \rangle}

\newcommand{\NL}{\text{NL}}

\begin{document}

\title{The preparation problem in nonlinear extensions of quantum theory}

\author{Eric G. Cavalcanti}
\affiliation{School of Physics, The University of Sydney, Sydney, NSW, 2006, Australia}
\affiliation{Centre for Quantum Dynamics, Griffith University, QLD 4111, Australia}

\author{Nicolas C. Menicucci}
\affiliation{School of Physics, The University of Sydney, Sydney, NSW, 2006, Australia}
\affiliation{Perimeter Institute for Theoretical Physics, 31 Caroline St N, Waterloo, ON, N2L 2Y5, Canada}

\author{Jacques L. Pienaar}
\affiliation{School of Mathematics and Physics, The University of Queensland, Brisbane, QLD, 4072, Australia}

\begin{abstract}
Nonlinear modifications to the laws of quantum mechanics have been proposed as a possible way to consistently describe information processing in the presence of closed timelike curves. These have recently generated controversy due to possible exotic information-theoretic effects, including breaking quantum cryptography and radically speeding up both classical and quantum computers. The physical interpretation of such theories, however, is still unclear. We consider a large class of operationally-defined theories that contain ``nonlinear boxes'' and show that operational verifiability without superluminal signaling implies a split in the equivalence classes of preparation procedures.  We conclude that any theory satisfying the above requirements  is (a)~inconsistent unless it contains distinct representations for the two different kinds of preparations and (b)~incomplete unless it also contains a rule for uniquely distinguishing them at the operational level. We refer to this as the {\it preparation problem} for nonlinear theories.  In addition to its foundational implications, this work shows that, in the presence of nonlinear quantum evolution, the security of quantum cryptography and the existence of other exotic effects remain open questions.

\end{abstract}


\maketitle


\textit{Introduction.}---Two outstanding problems in quantum mechanics are the problem of reconciling the quantum framework with the gravitational interaction and the problem of classical emergence from quantum dynamics. In both cases, it has been suggested that quantum mechanics might be an approximation to an underlying theory whose dynamics are nonlinear~\cite{DEU91,GRW,REG09,CHU12,PEN98,WEI89,BIA76,HAA78,GOL08,DOE92}. However, the consistent formulation and interpretation of such proposals for nonlinear extensions to quantum theory has proven problematic. In particular, such proposals run the risk of allowing the transmission of information faster than light~\cite{GIS90, JOR99} unless they are very carefully formulated~\cite{CZA98, KEN05}.

The question of how quantum systems evolve in the presence of gravity is at the heart of a model by Deutsch~\cite{DEU91} for describing their evolution near an extreme example of curved spacetime---that of a closed timelike curve~(CTC)~\cite{Godel1949, Morris1988, Gott1991}. In this toy model,  the quantum state of the system entering the CTC is identified with that of the system exiting the CTC (in the past). This constraint  leads to a nonlinear mapping between the input and output states. The discovery that this type of nonlinearity leads to exotic information-theoretic effects has spurred recent literature on the topic~\cite{BAC04, RAL07, BRU09, BEN09, RAL10, PIE11, BRU12}. Among these results, it was shown that  both quantum and classical computers with access to CTCs would become immensely more powerful~\cite{BAC04,BRU12,Aaronson2009} and that it is possible to distinguish non-orthogonal states~\cite{BRU09,BRU12} and thereby break quantum cryptographic protocols thought to be unconditionally secure~\cite{Bennett1984}. These results remain controversial, with some authors~\cite{BEN09} claiming that no advantage could be gained from Deutsch-model CTCs. Others disagree and have extended Deutsch's model to mixed-state inputs~\cite{RAL10} and quantum fields~\cite{PIE11}, using a formulation that would arguably allow these exotic effects to persist. Because these results have important foundational, physical, and technological implications, it is important to resolve this conflict. 

Recent work~\cite{CR11} suggests that no extension of quantum theory can have more predictive power than quantum theory itself. The theories we consider here, however,  do not satisfy the authors' assumption that any process can equally well be described by unitary evolution. This assumption also lies behind the argument for the impotence of Deutsch-model CTCs~\cite{BEN09}, and we address it in the discussion. Finally, we note that alternative models of CTCs exist~\cite{POL94,LLO11}. These will not be considered for reasons discussed below.

In this paper, we adopt an operational approach  to preparation procedures within a nonlinear framework, together with the assumption of no superluminal signaling (for compatibility with relativity). We  consider a class of nonlinear evolutions that are \textit{operationally verifiable} and show that they must give different  measurement statistics  for preparations that are in the same equivalence class under linear quantum mechanics. This raises the question of how to represent the new equivalence classes of preparations that arise under nonlinear evolution and how to determine which physical preparations belong to each class. Our formalism allows us to contrast and compare a broad class of non-signaling, nonlinear models according to their operational verifiability and the extent to which they resolve the ambiguities of state preparation.

\begin{figure}[!tbp]
 \includegraphics[width=8cm]{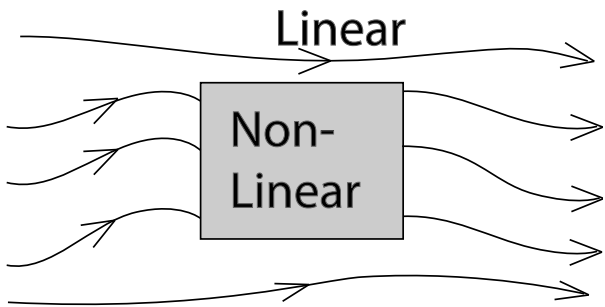}
\caption{A nonlinear box  is a bounded region of spacetime in which quantum states are mapped nonlinearly from inputs to outputs. Standard (linear) quantum mechanics is required to hold at the operational level everywhere outside the box, i.e.  we do not forbid objective collapse~\cite{GRW,PEN96} or other models of quantum theory consistent with experimental tests to date.}
\label{Fig_box}
\end{figure}

\textit{Nonlinear dynamics.}---Operationally,  a transformation on some known input states is called nonlinear if its output statistics cannot be replicated by standard linear quantum mechanics. A simple example  is Deutsch's toy model for quantum evolution in the presence of closed time-like curves (CTCs)~\cite{DEU91}. This model is not the most general form that a nonlinear evolution may take; however,  it has the desirable feature that the nonlinearity is confined to a finite region of spacetime outside of which the laws of physics obey standard quantum mechanics---i.e.,~the theory can be described as standard quantum mechanics augmented by a ``nonlinear box'' (Fig.~\ref{Fig_box}). 

The class of theories with nonlinear boxes includes, but is not limited to, any nonlinear theory that can be emulated using a ``state readout'' box~\cite{KEN05}, since these devices are explicitly confined to a local region of spacetime. There exist proposals for nonlinear models that do not appear to fit the description of a nonlinear box, such as the path-integral models for quantum evolution near CTCs~\cite{POL94} and their quantum-circuit counterparts (P-CTCs)~\cite{LLO11}. In those models,  the presence of nonlinearity in the CTC region places constraints on the laws of physics in distant regions, leading to highly nonlocal dynamics~\cite{CZA91,GIS90,Ralph2011}. Unless it can be shown that such theories are describable by nonlinear boxes, it is unclear how to represent them in an operational framework. We therefore restrict our attention to those models that explicitly conform to the nonlinear box paradigm. For emphasis and conceptual simplicity, we formulate our argument in terms of nonlinear evolution on  a set of pure states. These are operationally defined in standard quantum theory as states of maximal information, and this description is valid, by assumption, outside the nonlinear box.  Later on, we extend this analysis to more general nonlinear maps while remaining within the nonlinear box framework.

Consider a specific example in which the nonlinearity is strong enough to distinguish between a pair of non-orthogonal bases. Formally, we consider a box that implements the following nonlinear map:
\begin{align}
	\label{Brun} 
\ket{\psi_0} \otimes \ket{0} &\mapsto \ket{0} \otimes \ket{0} \,, & \ket{\phi_0} \otimes \ket{0} &\mapsto \ket{1} \otimes \ket{0} \,, \nonumber \\ 
\ket{\psi_1} \otimes \ket{0} &\mapsto \ket{0} \otimes \ket{1} \,, & \ket{\phi_1} \otimes \ket{0} &\mapsto \ket{1} \otimes \ket{1} \,,
\end{align}
where $\{ \ket{\psi_0}, \ket{\psi_1} \}$ and $\{ \ket{\phi_0}, \ket{\phi_1} \}$ are two non-identical orthogonal bases for the same qubit. (The computational basis---used for the second input qubit and for measurement---is labeled $\{ \ket{0}, \ket{1} \}$. It is arbitrary and could even be one of the other two bases.) Since the two bases are non-identical, the four states are not mutually orthogonal, and this map is forbidden by linear quantum theory as it permits non-orthogonal states to be perfectly distinguished. Note that the action of the nonlinear map on a basis of states does not in general specify its action on all states in the state space.

In evaluating the consequences of such exotic evolutions, we must exercise caution in selecting the correct formalism, for if we rely too heavily on the conventions of linear quantum mechanics we may run into contradictions~\cite{BEN09}. These can be resolved by an operational approach~\cite{SPE05}. Let us therefore recall just what is required by the statement that the evolution~\eqref{Brun} is found to occur.

The operational description of any theory can be phrased in terms of a physical system that undergoes a preparation~$P$ followed by a transformation~$T$ and finally a measurement~$M$ resulting in an outcome $k$~\cite{SPE05}. The operational theory assigns probability distributions $p(k|P,T,M)$ for outcome $k$ given some particular $P$, $T$, and~$M$. Once we have a rule for obtaining these probabilities, we can define \textit{equivalence classes} of preparations. Two preparation procedures $P$ and $P'$ are said to belong to the same equivalence class with respect to a given operational theory if they result in the same measurement statistics for all possible choices of $T$ and $M$. That is, $P$ and $P'$ are \textit{operationally equivalent} $(P\equiv P')$ if and only if $p(k|P,T,M)=p(k|P',T,M)$ for all $T$, $M$. Similar definitions hold for transformations and measurements, but we will not need them here. As an example from quantum mechanics, consider two possible preparations of the state $\ket{1}$. In the preparation $P^{(d)}$, a quantum NOT gate is applied to a qubit initially in the $\ket{0}$ state. In the preparation $P^{(e)}$, an ensemble of $\ket{+}$ states are all measured in the computational basis, before post-selecting one of them conditional on obtaining the outcome~``1". The distinct preparation procedures $P^{(d)}$ and $P^{(e)}$ are operationally equivalent under linear operations, but two preparations that are equivalent under linear operations may in general not be equivalent once we consider nonlinear transformations.

Our assumption that a given nonlinear map can be described as a nonlinear box means that the laws of physics outside the box are governed by standard operational quantum mechanics. Note that we do \textit{not} require that any specific interpretation of quantum mechanics holds, but just that the predictions of standard quantum mechanics in all currently testable regimes hold true far away from, and independently of, the nonlinear box. Then we may regard the nonlinear map \eqref{Brun} as defining a transformation $T_\NL$, where the inputs to the box are prepared by some physical preparations, and measurements are performed on the outputs. It follows that the particular nonlinear transformation~$T_\NL$ given by \eqref{Brun} is verifiable \textit{only if} there exists a set of preparations denoted $\mathcal{P}_V \coloneqq \{ P_V^{(\psi_0)},P_V^{(\psi_1)},P_V^{(\phi_0)},P_V^{(\phi_1)}\}$ that are equivalent under linear operations to preparations of the pure states $\{ \ket{\psi_0}, \ket{\psi_1}, \ket{\phi_0}, \ket{\phi_1} \}$, and a set of measurements $\mathcal{M}_V \coloneqq \{M_{V,j}:j\}$ such that the set of probabilities $p(M_{V,j}^{(k)}|P_V^{(i)},T_\NL,M_{V,j})$---where $k$ labels the measurement outcomes---are sufficient to identify the map~\eqref{Brun}. If the theory includes any caveats that make the set $\mathcal{P}_V$ or $\mathcal{M}_V$ empty, then the map~\eqref{Brun} is not verifiable. Notice that the non-emptiness of both $\mathcal{P}_V$ and $\mathcal{M}_V$ is merely a \textit{necessary} condition for verifiability. Some choices may not be sufficient because they would make a less radical explanation of the data available. We only need the necessity of this condition in order to reach our eventual conclusion.


\textit{Superluminal signaling.}---Assuming that \eqref{Brun} is verifiable, so that $\mathcal{P}_V$ and $\mathcal{M}_V$ are nonempty, we now apply a well-known ``remote preparation'' protocol that, absent additional restrictions, would enable us to send signals faster than light~\cite{GIS90}. Alice and Bob, each in possession of one half of the singlet state $\frac{1}{\sqrt{2}} \brac{ \ket{0_A 1_B} - \ket{1_A 0_B} } $, are separated by a great distance. Alice then performs a projective measurement on her half of the Bell pair. The measurement can be made in either the $\{ \ket{\psi_0}, \ket{\psi_1} \}$ basis or the $\{ \ket{\phi_0}, \ket{\phi_1} \}$ basis, at Alice's discretion. If we have access to Alice's measurement result, we can assign Bob's qubit a definite pure state immediately after the measurement. If we do not have access to this information, then we do not know which pure state has been prepared. (In ordinary quantum theory, we could simply assign an impure density matrix to this situation, but a priori, such is not the case for verifiable nonlinear evolution~\cite{Cavalcanti2010}.)

We note that regardless of how we choose to represent Bob's state in the formalism, this representation must be consistent with all future measurements performed upon the  system and future repetitions of this experiment. We therefore define the remote preparations operationally as follows: if Alice measures in the first basis, she implements either the preparation $P_R^{(\psi_0)}$ or $P_R^{(\psi_1)}$ (depending on the outcome), resulting in the operational state of Bob's qubit being $\ket{\psi_0}$ or $\ket{\psi_1}$ respectively. Specifically, we mean that, absent any interaction with the nonlinear box, future measurements on Bob's qubit will be consistent with one of these states. If instead Alice measures in the second basis, she implements one of the preparations $P_R^{(\phi_0)}$ or $P_R^{(\phi_1)}$, resulting in $\ket{\phi_0}$ or $\ket{\phi_1}$ for Bob's qubit. Let us denote this set of possible remote preparations by  $\mathcal{P}_R \coloneqq \{P_R^{(\psi_0)},P_R^{(\psi_1)},P_R^{(\phi_0)},P_R^{(\phi_1)}\}$. 

Suppose now that Bob possesses a nonlinear box that implements the transformations in Eqs.~(\ref{Brun}), provided he uses the preparations from $\mathcal{P}_V$ and measurements from $\mathcal{M}_V$. We observe that the remote preparations $\mathcal{P}_R$ and the verifying preparations $\mathcal{P}_V$ are operationally equivalent under linear quantum mechanics---i.e., $P_R^{(x)} \equiv P_V^{(x)}$, with $x \in \{ \psi_0, \psi_1, \phi_0, \phi_1 \}$. Let us therefore \textit{assume} that the remote preparations $\mathcal{P}_R$ belong to the set of preparations $\mathcal{P}_V$ that exhibit the evolution (\ref{Brun}). It then follows that Bob can distinguish the measurement basis chosen by Alice by sending the remotely prepared qubit through the nonlinear box and measuring the first output qubit, thus revealing which basis the input state belongs to. Under these assumptions, Alice may send information to Bob instantaneously. Since consistency with relativity forbids this, we must rethink our assumptions.


\textit{The preparation problem.}---Signaling can be avoided \textit{only if} the remote preparations in $\mathcal{P}_R$ are not in the same equivalence class as the corresponding preparations in $\mathcal{P}_V$ when nonlinear transformations are considered. As soon as we distinguish different preparations of the states $\{ \ket{\psi_0}, \ket{\psi_1}, \ket{\phi_0}, \ket{\phi_1} \}$, two problems immediately follow.

First, we see that the pure state of a physical system does not uniquely determine the system's evolution under the nonlinear map \eqref{Brun}, since only certain preparations $\mathcal{P}_V$ lead to the verification of this map, while others do not. We are then faced with the problem of how to represent quantum states in the formalism, given that the pure state alone is insufficient to determine the evolution. If we do not make a formal distinction between different preparations, then our model is inconsistent.

Second, while we have demonstrated that $\mathcal{P}_V$ cannot include all preparations of a given pure state (specifically, $\mathcal{P}_V$ cannot contain $\mathcal{P}_R$), we have not provided a strict rule for determining which physical preparations belong to $\mathcal{P}_V$  and which do not . Completeness of the model requires that the elements of $\mathcal{P}_V$ be unambiguously defined by such a rule.

These problems, collectively referred to as the \textit{preparation problem},  must be addressed by any verifiable, non-signaling model that admits evolutions of the form \eqref{Brun}.  While this formulation is not the most general one, it is sufficient to discuss the extent to which existing proposals for nonlinear theories in the literature address the preparation problem. A formulation in terms of more general nonlinear evolutions follows. 

\textit{Discussion.}---The evolution~\eqref{Brun} is  exhibited by the Deutsch model of CTCs~\cite{DEU91,BRU09} and its derivatives~\cite{BEN09,RAL10}, so these models are necessarily subject to the preparation problem.  While Deutsch's original paper~\cite{DEU91}  argued that signaling does not occur,  a complete discussion of different physical preparations is lacking, and therefore the problem remains open for this model.

Bennett \textit{el al.}~\cite{BEN09} attempted to fill this gap by arguing, in effect, that non-deterministic preparations (including a rational agent choosing by fiat) must be excluded from $\mathcal{P}_V$, and hence the strange effects predicted by the Deutsch model are  unverifiable by most (if not all) preparations. At face value, the authors' prescription is inconsistent: they start with the assumption that the CTC performs a nonlinear evolution like Eqs.~\eqref{Brun}, and yet Rob, who prepares one of two non-orthogonal pure states from the input set and presents it to Alice for discrimination, does not find the output of the CTC to be correlated with its input according to that map. This inconsistency may be lifted by giving Rob's state a different representation from those for which the nonlinear map holds. The model remains incomplete, however, as it does not specify what kinds of processes generate the states that \textit{do} evolve according to the nonlinear map.

The authors state that ``while in standard quantum mechanics the evolution of a mixture is equal to the corresponding mixture of the evolutions of the individual states, in a nonlinear theory, this is not generally true.'' This is a valid observation. But it does not follow that there cannot be certain preparations producing a preferred decomposition for which the mixing \textit{does} distribute over the individual evolutions. This could occur, for example, in an objective collapse theory~\cite{PEN96,GRW}. The authors of~\cite{BEN09} exclude this possibility by assumption, arguing that the output of \textit{any} process by which Rob chooses between $\ket{\phi_0}$ or $\ket{\phi_1}$ will be, for the purposes of the nonlinear evolution, given by the density matrix that represents that mixture in linear quantum theory. This assumption may hold within the so-called ``church of the larger Hilbert space,'' as the authors discuss, but it may not be true in a nonlinear extension of quantum theory.

An alternative to this view is to regard the Deutsch model as being represented by an equivalent circuit on an extended Hilbert space, as Ralph and Myers propose~\cite{RAL10}. This formalism maps the input state to a different state in the larger space depending on its preparation procedure. It is therefore consistent because it takes into account different physical preparations at the formal level. While all so-called ``remote'' preparations are declared excluded from $\mathcal{P}_V$, there is no unambiguous operational meaning to this. The states that can be used for verification are  those that are deterministically prepared by a suitably defined classical experimenter. But without a means of unambiguously distinguishing a ``classical experimenter'' from just another quantum system,  we must regard this model as incomplete.

Kent's ``state readout device''~\cite{KEN05}  is a nonlinear box that nondestructively outputs a classical record of the input system's density matrix.   Using this information, an experimenter can transform the input state into any other state, emulating a large class of nonlinear evolutions, including Eqs.~\eqref{Brun}.  The readout takes into account all classical data that exists about the state within the past light cone of the box, so that states prepared  from outside the light cone appear mixed, while those prepared entirely within the light cone are represented as pure. Since the readout of a state unambiguously determines whether or not it belongs in $\mathcal{P}_V$, the formalism is consistent, and since all states can be assigned a readout given their preparation procedure, the formalism is also complete. Both the Ralph-Myers and Kent models require a separation between the classical and the quantum (i.e.,~a ``Heisenberg cut''). The Ralph-Myers model is incomplete because no method is provided for unambiguously distinguishing between $\mathcal{P}_V$ and $\mathcal{P}_R$ at the operational level, while Kent's model uses the past light cone of the box for this purpose.

Finally, it remains to be seen whether ``completely separable'' nonlinear theories, as described by Czachor~\cite{CZA98}, can always be cast in terms of nonlinear boxes. If so, our results would apply to them, as well.

\textit{Generalization.}--- We  extend the class of theories subject to the preparation problem by considering generalized preparations instead of pure states and generalized measurements instead of projective ones, based on Gisin~\cite{GIS89}. Because we are working in the nonlinear box paradigm, these concepts inherit their operational significance unchanged from ordinary quantum theory.

Consider  a nonlinear map $\mathcal{N}$ acting on density operators. By definition, there must exist two sets of density operators $\{\rho_1^{(i)}:i\}$ and $\{\rho_2^{(j)}:j\}$ and probability measures $\{p_i:i\}$ and $\{q_j:j\}$ such that $\sum_i p_i \rho_1^{(i)}=\sum_j q_j \rho_2^{(j)}$ but where $\sum_i p_i \mathcal{N}(\rho_1^{(i)}) \neq \sum_j q_j \mathcal{N}(\rho_2^{(j)})$. Such a map takes what would be identical inputs in linear quantum theory to decomposition-dependent nonidentical outputs. The map \eqref{Brun} is one example.

Proceeding as before, $\mathcal{N}$~is operationally verifiable only if there exists a set of preparations $\mathcal{P}_V \mathrel{\coloneqq} \{ P_V^{(i)}:i\}$  for the states $\{\rho^{(i)}:i\}$ on which $\mathcal{N}$ is specified, and there exists a set of measurements $\mathcal{M}_V \mathrel{\coloneqq} \{M_{V,j}:j\}$ such that the probabilities $p(M_{V,j}^{(k)}|P_V^{(i)},T_\NL,M_{V,j})$---where $k$ labels the measurement outcomes---are sufficient to identify the map $\mathcal{N}$. The quantum steering theorem~\mbox{\cite{SCH36,GIS89,HJW93}}, as generalized by Verstraete~\cite{Ver02}, states that for any decomposition of a density matrix $\sigma_B = \sum_i p_i \rho^{(i)}$, there exists a state $\sigma_{AB}$ for a bipartite system such that $\mathrm{Tr_A}(\sigma_{AB})=\sigma_B$, and there exists a generalized measurement on $A$ that, when applied to $\sigma_{AB}$, will produce with probability $p_i$ the reduced state $\rho^{(i)}$ at $B$. Thus, for any nonlinear map $\mathcal{N}$, there exists a set of remote preparations $\mathcal{P}_R$ that can be used for signaling unless they are excluded from  $\mathcal{P}_V$, leading to the preparation problem. 

\textit{Conclusion.}---%
 Verifiable nonlinear evolution without superluminal signaling requires the splitting of normally equivalent classes of preparations  such that remote preparations cannot be used to verify the nonlinear evolution.  Any proposal for such a theory is therefore inconsistent unless it makes a formal distinction between these preparation procedures and incomplete unless it supplies an operational rule that determines which preparations can be used to verify the evolution and which cannot. Although we have  limited  discussion  to nonlinear-box-type models , other nonlinear models---for example, P-CTCs~\cite{LLO11}---may  suffer from similar problems.

This work has important implications for assessing the potential of nonlinear evolution to break quantum cryptographic schemes~\cite{BRU09,BRU12} and to perform other exotic feats~\cite{BAC04,BRU12,Aaronson2009}. In particular, it shows that recent arguments~\cite{BEN09}  against this possibility are questionable because they depend on arbitrary and untested assumptions, and the issue therefore remains open.  This work should also be seen in light of the program of studying theories generalizing quantum theory in search of principles that are sufficient to reconstruct it. Our study of nonlinear boxes is analogous to the study of nonlocal boxes~\cite{PR94}, which allow stronger-than-quantum correlations without signaling. Even if nonlinear evolution turns out to be unphysical, studying theories that contain nonlinear boxes may lead to insights into why quantum theory must be linear after all.


We thank Roger Colbeck, Daniel Gottesman, Tim Ralph, Todd Brun, Stephen Hawking, Jeffrey Bub, and Alexei Grinbaum for discussions. Research at Perimeter Institute is supported by the Government of Canada through Industry Canada and by the Province of Ontario through the Ministry of Research \& Innovation. E.G.C.\ received support from Australian Research Council Discovery grant DP0984863 and Discovery Early Career Researcher Award DE120100559. N.C.M.\ is supported by Australian Research Council Discovery Early Career Researcher Award DE120102204.

\bibliographystyle{bibstyle_notitle}
\bibliography{prep_prob}

\begin{thebibliography}{10}

\bibitem{DEU91}
D. Deutsch, Phys. Rev. D {\bf 44}, 3197 (1991).

\bibitem{GRW}
G.~C. Ghirardi, A. Rimini, and T. Weber, Phys. Rev. D {\bf 34}, 470 (1986).

\bibitem{REG09}
M. Reginatto and M.~J.~W. Hall, Journal of Physics: Conference Series {\bf
  174}, 012038 (2009).

\bibitem{CHU12}
A.~J.~K. Chua, M.~J.~W. Hall, and C.~M. Savage, Phys. Rev. A {\bf 85}, 022110
  (2012).

\bibitem{PEN98}
R. Penrose, Philosophical Transactions of the Royal Society of London. Series
  A: Mathematical, Physical and Engineering Sciences {\bf 356}, 1927 (1998).

\bibitem{WEI89}
S. Weinberg, Annals of Physics {\bf 194}, 336  (1989).

\bibitem{BIA76}
I. Bialynicki-Birula and J. Mycielski, Annals of Physics {\bf 100}, 62  (1976).

\bibitem{HAA78}
R. Haag and U. Bannier, Communications in Mathematical Physics {\bf 60}, 1
  (1978).

\bibitem{GOL08}
G. Goldin, Physics of Atomic Nuclei {\bf 71}, 884 (2008).

\bibitem{DOE92}
H.-D. Doebner and G.~A. Goldin, Physics Letters A {\bf 162}, 397  (1992).

\bibitem{GIS90}
N. Gisin, Physics Letters A {\bf 143}, 1  (1990).

\bibitem{JOR99}
T.~F. Jordan and Z.-E. Sariyianni, Physics Letters A {\bf 263}, 263  (1999).

\bibitem{CZA98}
M. Czachor, Phys. Rev. A {\bf 57}, 4122 (1998).

\bibitem{KEN05}
A. Kent, Phys. Rev. A {\bf 72}, 012108 (2005).

\bibitem{Godel1949}
K. G\"odel, Rev. Mod. Phys. {\bf 21}, 447 (1949).

\bibitem{Morris1988}
M.~S. Morris, K.~S. Thorne, and U. Yurtsever, Phys. Rev. Lett. {\bf 61}, 1446
  (1988).

\bibitem{Gott1991}
J.~R. Gott~III, Phys. Rev. Lett. {\bf 66}, 1126 (1991).

\bibitem{BAC04}
D. Bacon, Phys. Rev. A {\bf 70}, 032309 (2004).

\bibitem{RAL07}
T.~C. Ralph, Phys. Rev. A {\bf 76}, 012336 (2007).

\bibitem{BRU09}
T.~A. Brun, J. Harrington, and M.~M. Wilde, Phys. Rev. Lett. {\bf 102}, 210402
  (2009).

\bibitem{BEN09}
C.~H. Bennett {\it et~al.}, Phys. Rev. Lett. {\bf 103}, 170502 (2009).

\bibitem{RAL10}
T.~C. Ralph and C.~R. Myers, Phys. Rev. A {\bf 82}, 062330 (2010).

\bibitem{PIE11}
J.~L. Pienaar, C.~R. Myers, and T.~C. Ralph, Phys. Rev. A {\bf 84}, 062316
  (2011).

\bibitem{BRU12}
T. Brun and M. Wilde, Foundations of Physics {\bf 42}, 341 (2012).

\bibitem{Aaronson2009}
S. Aaronson and J. Watrous, Proc. Royal Soc. A {\bf 465}, 631 (2009).

\bibitem{Bennett1984}
C.~H. Bennett and G. Brassard, in {\em Proceedings of IEEE International
  Conference on Computers, Systems and Signal Processing, 1984}, \ pp.\
  175--179  (1984).

\bibitem{POL94}
H.~D. Politzer, Phys. Rev. D {\bf 49}, 3981 (1994).

\bibitem{LLO11}
S. Lloyd {\it et~al.}, Phys. Rev. Lett. {\bf 106}, 040403 (2011).

\bibitem{CR11}
R. Colbeck and R. Renner, Nat. Commun. {\bf 2}, 411 (2011).

\bibitem{PEN96}
R. Penrose, General Relativity and Gravitation {\bf 28}, 581 (1996).

\bibitem{CZA91}
M. Czachor, Foundations of Physics Letters {\bf 4}, 351 (1991).

\bibitem{Ralph2011}
T.~C. Ralph,  arXiv:1107.4675v1 [quant-ph] (2011).

\bibitem{SPE05}
R.~W. Spekkens, Phys. Rev. A {\bf 71}, 052108 (2005).

\bibitem{Cavalcanti2010}
E.~G. Cavalcanti and N.~C. Menicucci,  arXiv:1004.1219 [quant-ph] (2010).

\bibitem{GIS89}
N. Gisin, Helvetica Physica Acta {\bf 62}, 363 (1989).

\bibitem{SCH36}
E. Schr\"{o}dinger, Proc. Camb. Phil. Soc. {\bf 32}, 446 (1936).

\bibitem{HJW93}
L.~P. Hughston, R. Jozsa, and W.~K. Wootters, Phys. Lett. A {\bf 183}, 14
  (1993).

\bibitem{Ver02}
F. Verstraete, Ph.D. thesis, Katholieke Universiteit Leuven, 2002.

\bibitem{PR94}
S. Popescu and D. Rohrlich, Foundations of Physics {\bf 24}, 379 (1994).

\end{thebibliography}

\end{document}